\documentclass{spie} 

\usepackage{amsmath,amsfonts,amssymb}
\usepackage{graphicx}
\usepackage{tikz}
\newcommand{\argmin}{\operatornamewithlimits{argmin}}

\usepackage{lipsum}
\usepackage{multirow}
\usepackage{rotating}

\title{Randomness and isometries in echo\\ state networks and compressed sensing}
\author{Ashley Prater-Bennette \\ Air Force Research Laboratory \\  Rome NY 13441 USA}

\authorinfo{ashley.prater.3@us.af.mil.   Cleared for public release by WPAFB Public Affairs on 30 Jan 2018, case number 88ABW-2018-0407.  
Any opinions, findings and conclusions or recommendations expressed in this work are those of the author and do not necessarily reflect the view of the United States Air Force. 
}

\begin{document} 
\maketitle

\begin{abstract}
Although largely different concepts, echo state networks and compressed sensing models both rely on collections of random weights; as the reservoir dynamics for echo state networks, and the sensing coefficients in compressed sensing. Several methods for generating the random matrices and metrics to indicate desirable performance are well-studied in compressed sensing, but less so for echo state networks.  This work explores any overlap in these compressed sensing methods and metrics for application to echo state networks.  Several methods for generating the random reservoir weights are considered, and a new metric, inspired by the restricted isometry property for compressed sensing, is proposed for echo state networks.    The methods and metrics are investigated theoretically and experimentally, with results suggesting that the same types of random matrices work well for both echo state network and compressed sensing scenarios, and that echo state network classification accuracy is improved when the proposed restricted isometry-like constants are close to 1.

\end{abstract}


\section{INTRODUCTION}
\label{sec:intro}  

This paper considers similarities between the random matrices used in compressed sensing and echo state networks (ESNs), as well as metrics that may be used in both concepts to predict good performance.  

Compressed sensing~\cite{candes_TIT, candes_SPM, donoho} is concerned with the recovery of a sparse vector $\beta \in \mathbb{R}^{N}$ given only the collection 
\begin{equation}\label{eq:CS}
	y = W\beta + \varepsilon
\end{equation}
of noisy linear measurements and information about the sensing matrix $W\in\mathbb{R}^{M\times N}$.  Generally in compressed sensing, $M\ll N$, yielding an overdetermined system in Equation~\eqref{eq:CS} that admits infinitely many solutions.  However, if $\beta$ is sufficiently sparse and $W$ satisfies certain properties, then the unique solution of~\eqref{eq:CS} may be approximated by a convex optimization scheme~\cite{chen, donoho}
\begin{equation*}
	\widehat{\beta} = \argmin_{\beta \in \mathbb{R}^N} \left\{\; \left\| \beta \right\|_1 \; : \; \left\| W\beta - y\right\|_\infty \leq \delta \;\right\}.
\end{equation*}
Several strategies for setting the sensing matrix $W$ have been used, including both random~\cite{candes_SPM, candes_TIT, donoho} and deterministic~\cite{arash, calderbank} methods. The common thread among all the strategies is that $W$ must satisfy a~\emph{restricted isometry property}, to be further explored in Section~\ref{sec:measures}, to guarantee the recoverability of $\beta$ from the observations~\eqref{eq:CS} with high probability.

On the other hand, ESNs are special types of recurrent neural networks whose hidden layer weights are not trained for a particular dataset or task, but are fixed.  The weights may be deterministic, as in delay line reservoirs~\cite{paquot, larger}, or randomly assigned as in classical ESNs~\cite{bertschinger, jaeger, maass} .  In this work, only classical ESNs are considered.  Properties of the randomly assigned hidden layer weights $W$ that contribute to good performance of ESNs are not fully understood. 
Generally one scales the matrix $W$ of hidden layer weights so that the spectral radius satisfies $|W|_* < 1$ in order to achieve favorable reservoir dynamics~\cite{jaeger, luko}, however this scaling is not a necessary nor sufficient condition for good performance.

This paper will explore the strategies and metrics designed for random matrices in each field of research, and apply them toward the other.  The methods and metrics are explored both theoretically in Sections~\ref{sec:measures} and~\ref{sec:weights}, and experimentally in Section~\ref{sec:experiments}.  A new metric to suggest desirable performance of an ESN based only on the random weight matrix, inspired by the RIP for compressed sensing, is proposed in Section~\ref{sec:measures}.

This work uses the following notation.  Matrices are denoted by capital Roman letters, e.g.\ $X$, and vectors by lowercase Roman letters, $x$.  Matrices will use `matlab' notation, where the {j-th} row of the matrix $X$ is given by $X(j,:)$, and the {k-th} column by $X(:,k)$.  The j-th element in a vector will be given by either subscripts of function notation: $x_j$ or $x(j)$.  Scalars will be denoted by lowercase Greek letters, e.g. $\alpha$, except for indices of arrays which will be denoted by $j, k$ or $t$. Throughout, $\|\cdot\|_p$ is the standard vector p-norm, and $|\cdot|_*$ represents the spectral radius, i.e.\ the largest absolute eigenvalue of the square matrix argument.

\section{ECHO STATE NETWORKS}
\label{sec:ESN}

In this section, the dynamics of ESNs are further explored.  
Suppose the spatiotemporal input signals are given by a matrix $A \in \mathbb{R}^{L \times T}$, where $L$ is the spatial dimension and $T$ is the temporal dimension of the inputs.   The hidden layer, sometimes also referred to as the \emph{reservoir}, has $N$ nodes.  The values of the nodes at all timesteps are stored in the matrix $X\in\mathbb{R}^{N\times T}$, and are updated according to the dynamics
\begin{equation}\label{eq:ESN}
	X(:,t+1) = (1-\alpha)X(:,t) + \alpha f \left( \beta W_\text{in} A(:,t) + \rho W X(:,t) + \gamma \right),
\end{equation}
where $\alpha \in [0,1]$ is the leaking rate, $\beta, \rho > 0$ are scaling factors, $\gamma$ is a bias term, and $f$ is a nonlinear activation function.  The matrices $W_\text{in}\in\mathbb{R}^{N\times L}$ and $W\in\mathbb{R}^{N\times N}$ are the input and reservoir (or hidden layer) weights, respectively. 
The hidden layer is fixed, but the output layer is trained for the specific dataset and desired task.    The output weights $W_\text{out}\in\mathbb{R}^{K\times N}$ are found so that the collection of input and desired output pairs $\{(A,Y)\}$ in the training set, $W_\text{out}A \approx Y$.  
This approximation is usually obtained using a regularization, such as 
\begin{equation*}
	W_\text{out} = \argmin_{W\in\mathbb{R}^{K\times N}} \left\{ \left\| Y - WA\right\|^2_2 + \lambda \left\|W\right\|_2^2 \right\},
\end{equation*}
with approximate solution~\cite{tibshirani, chen}
\begin{equation}\label{eq:Wout}
	W_\text{out} \approx YX^\top \left( XX^\top + \lambda I_{N\times N}\right)^{-1},
\end{equation}
for the regularization constant $\lambda>0$.
Note that a \emph{single} output weight matrix is found and applied to the hidden layer states at all time steps.  Alternate approaches have been proposed, but for linear output weights with matrices, the approach used in this work tends to yield the most accurate output results~\cite{Prater_IJCNN}.

\begin{figure}[hptb]
	\centering
	\begin{tikzpicture}

		\draw (1,1) -- (1.5,1) -- (1.5,6) -- (1,6) -- (1,1);
		\draw (1.25,6) node[anchor=south] {Input};
		\draw (1.25,1) node[anchor=north] {$A(:,t)$};

		\draw (11,1.5) -- (10.5,1.5) -- (10.5,5.5) -- (11,5.5) -- (11,1.5);
		\draw (10.75,6) node[anchor=south] {Output};
		\draw (10.75,1) node[anchor=north] {$Y(:,t)$};

		\draw (6,3.5) circle (2.5) ;
		\draw (6,6) node[anchor=south] {Hidden Layer};
		\draw (6,1) node[anchor=north] {$X(:,t)$};

		\node (N4) at (4.02, 4.49) {$n_0$};
		\node (N6) at (4.02, 2.50) {$n_1$};
		\node (N9) at (5.07, 3.82) {$n_2$};
		\node (N3) at (5.45, 5.70) {$n_3$};
		\node (N8) at (5.55, 1.25) {$n_4$};
		\node (N7) at (6.13, 2.33) {$n_5$};
		\node (N5) at (6.67, 4.28) {$n_6$};
		\node (N0) at (7.40, 1.72) {$n_7$};
		\node (N2) at (7.43, 5.29) {$n_8$};
		\node (N1) at (8.30, 3.45) {$n_9$};

		\path[->] (N1) edge node {} (N3) ;
		\path[->] (N2) edge node {} (N1);
		\path[->] (N2) edge node {} (N3);
		\path[->] (N2) edge node {} (N7);
		\path[->] (N2) edge node {} (N9);
		\path[->] (N2) edge node {} (N0);
		\path[->] (N3) edge node {} (N9);
		\path[->] (N4) edge node {} (N9);
		\path[->] (N5) edge node {} (N7);
		\path[->] (N5) edge node {} (N6);
		\path[->] (N6) edge node {} (N8);
		\path[->] (N6) edge node {} (N9);
		\path[->] (N7) edge node {} (N0);
		\path[->] (N7) edge [loop left] ();
		\path[->] (N8) edge node {} (N6);
		\path[->] (N0) edge node {} (N3);

		\path[->] (1.6,3.5) edge node[anchor=south] {$W_\text{in}$} (3.4,3.5);
		\path[->] (8.6,3.5) edge node[anchor=south] {$W_\text{out}$} (10.4,3.5);
	\end{tikzpicture}
	\caption{A representation of an ESN with $9$ hidden layer nodes.}
	\label{fig:ESN}
\end{figure}
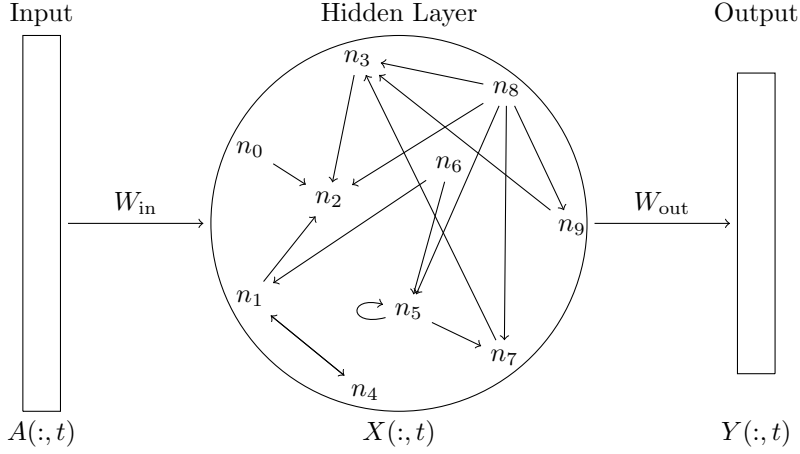

ESNs are used for both classification tasks and temporal prediction tasks.  In temporal prediction tasks, the desired output is a `future input', i.e $K=L$ and $Y(:,t) = A(:,t+d)$ for some temporal offset $d>0$.  In classification tasks, the desired output is an incidator of current class membership.  That is, let $K$ equal the number of classes, and choose $Y(:,t) = e_k$, a column vector of all zeros except for a 1 in the $k$-th row, if the corresponding training input belongs to the $k$-th class at time $t$.

%
\section{QUALITY MEASURES}\label{sec:measures}
It is desirable to determine metrics that can be used to indicate whether a particular instance of a random matrix will perform well in an ESN or compressed sensing model before deploying it on a particular dataset.  In this section, two types of measures of the quality are discussed, namely how closely the random dynamics  represent an isometry, and how well clases are separated in the hidden layer of an ESN.  

\subsection{ISOMETRIC PROPERTIES}
Our study of the isometric behavior of the hidden layer of an ESN is inspired by the so-called \emph{restricted isometry property} (RIP) that may be exhibited by random matrices that arise in compressed sensing research~\cite{candes_TIT, candes_SPM}.  The RIP is used in compressed sensing to help indicate when a sparse vector $x$ may be recovered from the observations
${y = W\beta + \varepsilon}$ as in Equation~\eqref{eq:CS} using a low-rank recovery scheme.  The matrix $W\in\mathbb{R}^{M \times N}$ is typically a `short-and-fat' random matrix, with $M\ll N$.   Therefore Equation~\eqref{eq:CS} should in general be difficult to solve, since $B$ defines an overdetermined system.  It is both the sparsity of $\beta$ \emph{as well as the properties of $W$} than enable exact recovery. Say that $W$ satisfies the RIP if there exist some small constants $\delta_\ell$, $\delta_r > 0$ such that
\begin{equation}\label{eq:RIP CS}
	\left( 1 - \delta_\ell\right) \left\| \beta \right\|_2 \leq \left\| W_S \beta \right\|_2 \leq \left(1+\delta_r\right) \left\|\beta\right\|_2,
\end{equation}
for all $S$-sparse vectors $\beta$ and for all submatrices $W_S$ formed by taking $S$ columns from $W$.    Define the \emph{restricted-isometry interval} of a random matrix $W$ as
\begin{equation*}
	\mathrm{RII}_{W} := [1-\delta_\ell, 1+\delta_r],
\end{equation*}
where $\delta_\ell$ and $\delta_r$ are minimal over all such values that satisfy~\eqref{eq:RIP CS}.

Of course, ESNs are a completely different setting from compressed sensing.  With ESNs, one is not interested in recovering a sparse vector, but rather to embed inputs into a higher dimensional space using a rich set of dynamics in the reservoir.  Moreover, the hidden layer weights matrix in an ESN are square, and not concerned with sparsity of the inputs.  However isometry-like properties can still influence the quality of the reservoir.  In this setting, we propose the \emph{near-isometry property}, which may be expressed as
\begin{equation}\label{eq:RIP ESN}
	a \left\|x\right\|_2 \leq \left\|\rho W x \right\|_2 \leq b \left\| x \right\|_2, \quad \forall x \in \mathbb{R}^N
\end{equation}
where $a<b$ are real constants, and $\rho$ is as in Equation~\eqref{eq:ESN}.  Define the \emph{near-isometry interval} of a random matrix $W$ with scaling factor $\rho$ as
\begin{equation*}
	\mathrm{NII}_{\rho W} := [a,b],
\end{equation*}
where $a$ and $b$ are chosen as the tightest bounds that satisfy~\eqref{eq:RIP ESN}. 
It is intuitive that a larger near-isometric interval $\mathrm{NII}_{\rho W}$ would lead to better separation of data within the reservoir.  
However, if the near-isometry constants $a$ and $b$ are too large, then the reservoir may be saturated and give poor results.

Note that the restricted isometry and near-isometry intervals depend only on $W$ and $\rho$, and are agnostic to the particular input data of interest.

\subsection{SEPARATION}
For classification tasks in an ESN, it may be useful to measure how well the reservoir separates classes.  
The separation ratio~\cite{goodman, gibbons} measures the separation of classes in the reservoir at each time step.  Define the \emph{center of mass} of the reservoir nodes at time $t$ for inputs in the $k$-th class as
\begin{equation*}
	M_k(t) = \frac{1}{\left| \mathcal{C}_k\right|} \sum \left\{ X(:,t) ; \text{ for inputs $A$ in the $k$-th class $\mathcal{C}_k$}\right\}.
\end{equation*}
That is, $M_k(t)\in\mathbb{R}^{N\times 1}$ is the average value of the reservoir at time $t$ for all inputs in the $k$-th class of the training set.
The \emph{inter-class distance} is then defined as the mean distance between paris of class means at each time step:
\begin{equation*}
	d(t) = \frac{1}{K^2} \sum_{j=1}^K \sum_{k=1}^K \left\| M_k(t) - M_j(t) \right\|_2.
\end{equation*}
Finally, the \emph{intra-class variance} is the mean variance of the reservoir states within each class at each time step:
\begin{equation*}
	v(t) = \frac{1}{K} \sum_{k=1}^K \frac{1}{\left| \mathcal{C}_k\right|} \sum \left\{ \left\| M_k(t) - X(:,t)\right\|_2 \; : \; \text{ for inputs $A \in \mathcal{C}_k$ }\right\}.
\end{equation*}
The \emph{separation ratio} of the ESN at time $t$ is then defined as:
\begin{equation*}
	\mathrm{Sep}(t) = \frac{d(t)}{1 + v(t)}.
\end{equation*}

Intuitively, if the separation ratio is larger, then classes should be better separated in the reservoir and therefore lead to better output classification accuracy.  

\section{MATRICES OF RANDOM WEIGHTS}\label{sec:weights}
Implementations of ESNs or compressed sensing models often use a collection of random weights.  However, these randomizations may be achieved in several ways, some of which may give better results than others.  We will explore five methods for determining these random matrices $W\in\mathbb{R}^{M\times N}$: 
\begin{enumerate}
	\item[(M1)] Sample the columns of $W$ uniformly at random on the $N$-dimensional unit sphere $\mathbb{S}^{N-1}$.
	\item[(M2)] Sample the entries of $W$ i.i.d.\ from the normal distribution with mean 0 and variance $1/M$.
	\item[(M3)] Sample the entries of $W$ i.i.d.\ from the uniform distribution on $[-1,1]$.
	\item[(M4)] Sample $S<M$ entries of $W$ i.i.d.\ from the uniform distribution on $[-1,1]$, and set the remaining entries to $0$.
	\item[(M5)] Sample $S<M$ entries of $W$ i.i.d.\ from the normal distribution with mean 0 and variance $1/M$, and set the remaining entries to $0$.
\end{enumerate}

Methods (M1) and (M2) are often used to generate the sensing matrices in compressed sensing~\cite{candes_SPM, candes_TIT}.  Methods (M3)-(M5) may be found as strategies in ESN literature~\cite{jaeger, luko, Prater_IJCNN, prater_NN}.  Matrices $W\in\mathbb{R}^{M\times N}$ generating using (M1) or (M2) with $\rho = 1$ satisfy the RIP in Equation~\eqref{eq:RIP CS} with high probability provided ${M \geq cS\log(N/S)}$ for some constant $c>0$\cite{candes_SPM, candes_TIT}, where $S$ is the sparsity of the input signal.  Since $S=M=N$ in an ESN, this condition reduces to $W$ generated via (M1) or (M2) satisfying the near-isometry property~\eqref{eq:RIP ESN} for any $N$ for some $a$ and $b$ close to 1.

\begin{figure}[hptb]
\centering
\includegraphics[width=0.45\textwidth]{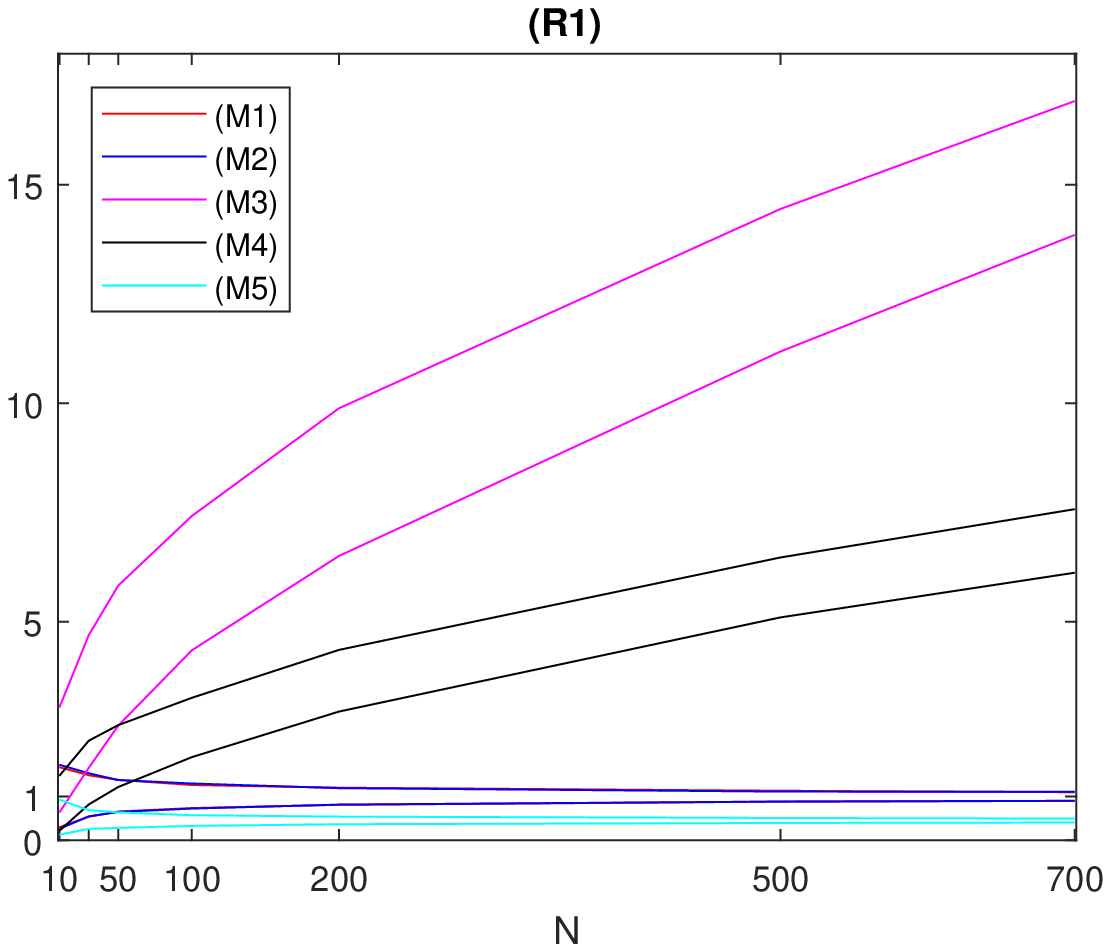}
\includegraphics[width=0.45\textwidth]{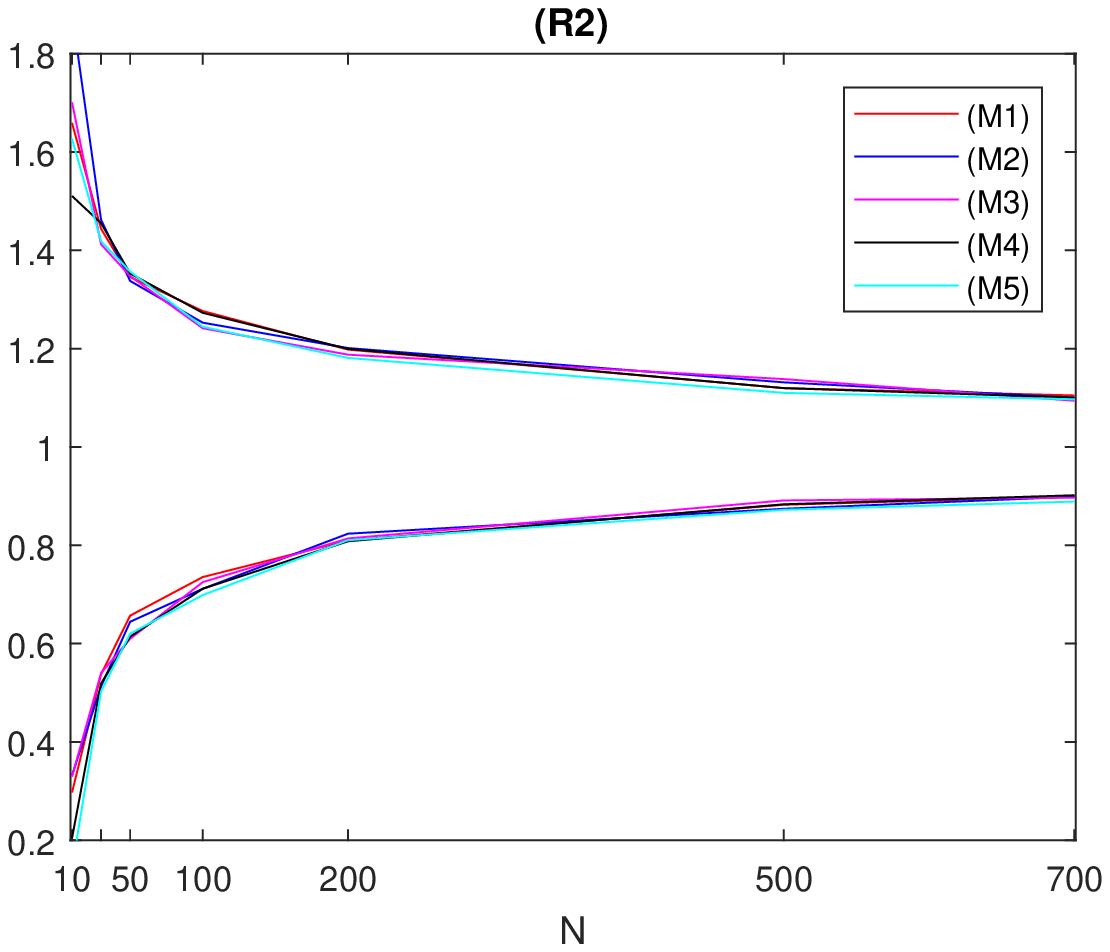}
\caption{Estimates for the upper and lower bounds $a$ and $b$ of the near-isometry interval for $\rho W\in\mathbb{R}^{N\times N}$ plotted against various $N$.  The colored lines correspond to different methods (M1)-(M5) to determine $W$.  The scaling factor $\rho$ is determined using (R1) in the left plots, and (R2) in the right plots. }
\label{fig:Interval}
\end{figure}

The choice of scaling factor $\rho$ will have a large impact on the behavior of the random dynamics in an ESN.  As shown in Figure~\ref{fig:Interval}, the scaling factor $\rho$ can be chosen so the matrices generated using methods (M3)-(M5) exhibit similar behavior as (M1) or (M2).  The figure displays the lower and upper bounds of the near-isometry intervals for reservoirs of size $n$ generated using methods (M1)-(M5).  For each color, the lower curve are values of $b$, and the upper curve are values of $a$.  In the left plot, $\rho$ is chosen to be $1$ throughout.  In the right plot, $\rho$ is chosen to be the inverse of the median value of the near-isometry interval, determined experimentally.   In addition to the two strategies discussed above, we will consider three other methods for determining $\rho$:
\begin{enumerate}
	\item[(R1)] Set $\rho = 1$.
	\item[(R2)] Set $\rho = 2/(a+b)$, where $[a,b]$ is the near-isometry interval computed on a sample set.
	\item[(R3)] Set $\rho = 1/\|W|_*$. 
	\item[(R4)] Set $\rho = 0.9/|W|_*$.
	\item[(R5)] Set $\rho = 1/ \|W\|_2$, where $\|\cdot\|_2$ is the matrix 2-norm, equalling the largest singular value.
\end{enumerate}

\section{EXPERIMENTAL RESULTS}\label{sec:experiments}
In this section, the random matrix methods discussed in Section~\ref{sec:weights} and quality measures discussed in Section~\ref{sec:measures} are explored experimentally when used in both echo state networks and compressed sensing models.  The random matrices are used as the reservoir weights in an ESN for a classification task, and used as the sensing matrix in a compressed sensing task.  Along with output accuracy for both types of tasks, the near-isometry interval and separation ratio are measured for ESNs, and the restricted isometry constants are measured for compressed sensing.


\subsection{ESN SETUP}
\begin{figure}[hptb]
\centering
\includegraphics[width=0.95\textwidth]{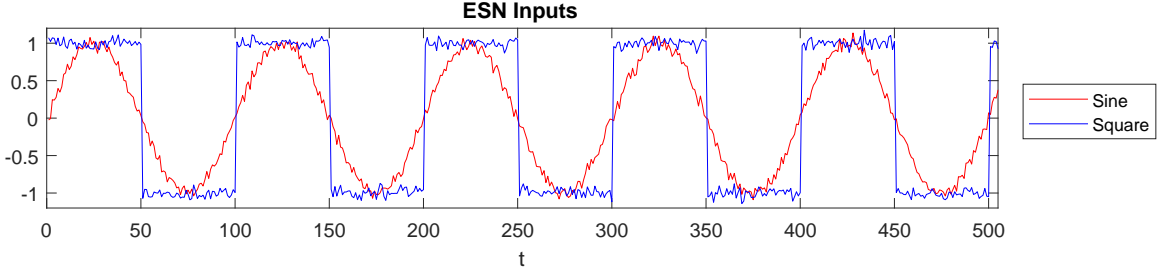}
\caption{Typical sine vs.\ square wave ESN inputs.}
\label{fig:Inputs}
\end{figure}

The simulations perform classification on a noisy sine versus square wave dataset.  The data consist of sine and square wave segments, each of period 100, repeated 150 times, then contaminated with gaussian noise $\sim N(0,0.05)$.   An illustration of the inputs is shown in Figure~\ref{fig:Inputs}.  The two types of signals are input sequentially into an ESN, which is set up using one of methods $(M1)-(M5)$ to generate $W$ with $N=200$ nodes, and one of (R1)-(R5) for the scaling factor.  The input weights $W_\text{in} \in \mathbb{R}^{L\times N}$ are chosen to be a matrix of all ones.  Other parameters from the ESN model~\eqref{eq:ESN} are chosen as ${\alpha = 1}$, ${f = \tanh}$, ${\beta = 1}$, ${\gamma = \pi/4}$, ${L=1}$, and ${K=2}$.  For methods $(M4)$ and $(M5)$, $S$ is chosen so only 20\% of the entries in $W$ are nonzero.

To train the output weights, Equation~\eqref{eq:Wout} is used where $X\in\mathbb{R}^{200\times 300,000}$ is the horizontal concatenation of the reservoir responses of all of the inputs, and the columns of the target outputs $Y\in\mathbb{R}^{2\times 300,000}$ equal $\begin{bmatrix} 1 & 0\end{bmatrix}^\top$ if the corresponding column of $X$ is generated by a sine wave input, and equal $\begin{bmatrix} 0 & 1 \end{bmatrix}^\top$ if generated by a square wave input.  For each pair $(Mj, Rk)$, the simulations are repeated 20 times, with new randomizations in $W$ and noise generated for each simulation.

\subsection{ESN RESULTS}
The results of the simulations are summarized in Table~\ref{tab:results}.  
Several measures of the quality of the ESN and the results are given.  `Acc' is the overall percent accuracy of the pointwise classification of the inputs, and `Acc Sine', `Acc Square' are the percent pointwise classification accuracy on the two types of inputs.  The spectral radius of the reservoir with scaling is measured by $|\rho W|_*$. The average separation ratio of the reservoir over all $t$ is given by $\overline{\mathrm{Sep}}$, and $[a,b]$ in the final column is the approximate near-isomorphism interval as in Equation~\eqref{eq:RIP ESN}.  Each entry is the average over 20 simulations.

The near-isomorphism intervals cannot be measured exactly, so their endpoints are estimated by 
\begin{equation}\label{eq:RIP est}
	a \approx \min\frac{\left\| W x \right\|_2}{\| x\|_2}, \quad b \approx \max \frac{\left\| W x\right\|_2}{\|x\|_2},
\end{equation}
where the extrema are taken over 10,000 instances of $x$ generated with entries sampled from the uniform distribution on $[-1,1]$. 


%
%
\begin{table}[htpb]
\centering
\begin{tabular}{c c || c | c | c | c | c | c | c }
\multicolumn{2}{c||}{Methods} 	&$\rho$	&Acc	&Acc Sine	&Acc Square	&$|\rho W_\text{res}|_*$	&$\overline{\mathrm{Sep}}$ 	&$[a,b]$ \\
\hline \hline
M1	&R1	& 1.00  &{\color{red}93.42}   &87.08 	&99.75 	 &1.03 	&{\color{red}4.28} 	&{\color{red}$[0.81, 1.19]$}\\
M1	&R2 	& 0.99  &{\color{red}93.51}   &87.27 	&99.75 	 &1.03 	&{\color{red}4.28}		&{\color{red}$[0.81, 1.19]$}\\
M1	&R3	& 0.95  &93.56   &87.42 	&99.71 	 &1.00 	&4.05 	&$[0.77, 1.15]$\\
M1	&R4	& 0.86  &93.19   &86.92 	&99.45 	 &0.90 	&3.75 	&$[0.70, 1.04]$\\
M1	&R5	& 0.51  &87.36   &76.12 	&98.60 	 &0.53 	&2.99		&$[0.41, 0.61]$\\
\hline
M2	&R1	& 1.00  &{\color{red}93.50}   &87.27 	&99.73 	 &1.05 	&{\color{red}4.21} 	&{\color{red}$[0.82, 1.19]$}\\
M2	&R2	& 0.99  &{\color{red}93.31}    &86.91 	&99.71 	 &1.04 	&{\color{red}4.32}		&{\color{red}$[0.81, 1.19]$}\\
M2	&R3	& 0.95  &93.42   &87.18 	&99.65 	 &1.00 	&4.14 	&$[0.77, 1.14]$\\
M2	&R4	& 0.87  &93.37   &87.29 	&99.45 	 &0.90 	&3.77 	&$[0.70, 1.03]$\\
M2	&R5	& 0.51  &87.77   &76.91 	&98.63 	 &0.52 	&3.02 	&$[0.41, 0.61]$\\
\hline
M3	&R1	& 1.00  &53.96   &55.84 	&52.08 	 &8.48 	&1.74		&$[6.66, 9.76]$\\
M3	&R2	& 0.12  &{\color{red}93.35}    &86.98 	&99.71 	 &1.04 	&{\color{red}4.16} 	&{\color{red}$[0.81, 1.20]$}\\
M3	&R3	& 0.12  &{\color{red}93.43}   &87.19 	&99.67 	 &1.00 	&{\color{red}4.16}		&{\color{red}$[0.79, 1.16]$}\\
M3	&R4	& 0.11  &93.31   &87.19 	&99.43 	 &0.90 	&3.76 	&$[0.71, 1.04]$\\
M3	&R5	& 0.06  &87.65   &76.74 	&98.56 	 &0.53 	&3.01		&$[0.41, 0.61]$\\
\hline
M4	&R1	& 1.00  &67.65   &66.42 	&68.88 	 &3.78 	&2.34 	&$[2.96, 4.39]$\\
M4	&R2	& 0.27  &{\color{red}93.46}    &87.21 	&99.71 	 &1.04 	&{\color{red}4.37} 	&{\color{red}$[0.81, 1.20]$}\\
M4	&R3	& 0.27  &{\color{red}93.52}   &87.29 	&99.75 	 &1.00 	&{\color{red}4.21} 	&{\color{red}$[0.78, 1.17]$}\\
M4	&R4	& 0.24  &93.29   &87.11 	&99.47 	 &0.90 	&3.79 	&$[0.70, 1.03]$\\
M4	&R5	& 0.14  &87.38   &75.98 	&98.78 	 &0.52 	&3.02 	&$[0.40, 0.60]$\\
\hline
M5	&R1	& 1.00  &85.74   &72.42 	&99.06 	 &0.46 	&2.94 	&$[0.36, 0.54]$\\
M5	&R2	& 2.22  &{\color{red}93.20}    &86.76 	&99.64 	 &1.04 	&{\color{red}4.18} 	&{\color{red}$[0.81, 1.19]$}\\
M5	&R3	& 2.15  &{\color{red}93.37}   &87.11 	&99.63 	 &1.00 	&{\color{red}4.10} 	&{\color{red}$[0.78, 1.16]$}\\
M5	&R4	& 1.92  &93.28   &87.19 	&99.38 	 &0.90 	&3.69 	&$[0.70, 1.03]$\\
M5	&R5	& 1.11  &87.06   &75.20 	&98.92 	 &0.51 	&2.98 	&$[0.40, 0.59]$\\
\end{tabular}
\caption{Results of the classification simulations using an ESN on the sine vs. square wave dataset.  The first two columns indicate the strategy used to generate $W_\text{res}$ and $\rho$, with the remaining columns giving the mean of several measures of quality over 20 simulations. The entries highlighted in red correspond to those with the highest separation and most symmetric isometry intervals close to 1, for each (Mj). }
\label{tab:results}
\end{table}

Good classification accuracy may be achieved using any method to generate $W$.  No one single method (Mj) appears to have an advantage over the others for this task.  However, the scaling factor $\rho$ has a large impact on the accuracy results.  Across all (Mj), selecting $\rho$ as in method (R2) consistently gives good accuracy, high class separation ratio, and tight near-isometry interval centered about 1.   Note that choosing $\rho$ so that $|\rho W|_* < 1$ as in method (R4) gives decent, but not the best, accuracy results across all (Mj), and also gives less favorable near-isometry intervals.  This approach is often used in ESN experiments to achieve good results.  These experimental results suggest that a better approach would be to optimize the near-isometry interval instead.

\subsection{COMPRESSED SENSING SETUP}
Sparse vectors are generated then recovered from noisy random linear observations.  The sparse vectors $\beta\in\mathbb{R}^{800}$ are generated by randomly selecting a set $S$ of $30$ indices, then sampling $\beta_{S(i)} = \gamma_i (1 + |a_i|)$, where $\{\gamma_i\}$ are i.i.d.\ randm variables sampled from the uniform distribution on $\{-1,1\}$ and $\{a_i\}$ are i.i.d.\ random variables sampled from the standard normal distribution.  For $i\notin S$, $\beta_i=0$.  The sparse vectors $\beta$ are observed according to
\begin{equation*}
	y = \rho W\beta + \varepsilon,
\end{equation*}
where the sensing matrix $W\in\mathbb{R}^{200 \times 800}$ is generated using methods (M1)-(M5), the scaling factor $\rho$ is generated using methods (R1), (R2) or (R5), and gaussian white noise $\varepsilon$ sampled i.i.d. $\sim N(0,0.05)$.   Note that the methods (R3) and (R4) are not used to generate $\rho$ because they do not apply to non-square matrices.

For each pair (Mj,Rk), 20 simulations are performed with new randomizations in $W, \beta, \varepsilon$ for each simulation.
The recovered sparse vector $\widehat{\beta}$ is found using the Dantzig selector method~\cite{candes_dantzig, prater_csda}.

\subsection{COMPRESSED SENSING RESULTS}
\begin{figure}
\centering
\includegraphics[width=0.48\textwidth]{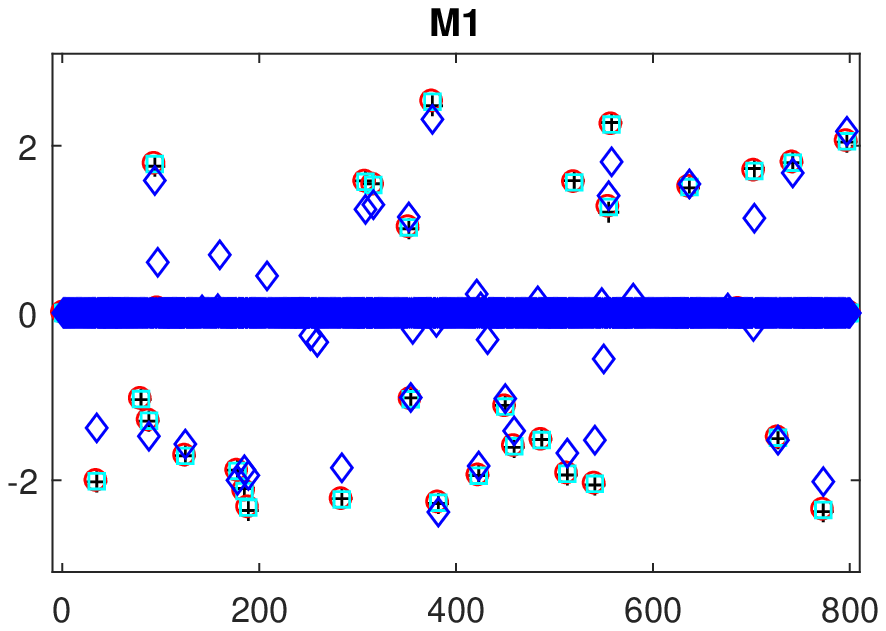}
\includegraphics[width=0.48\textwidth]{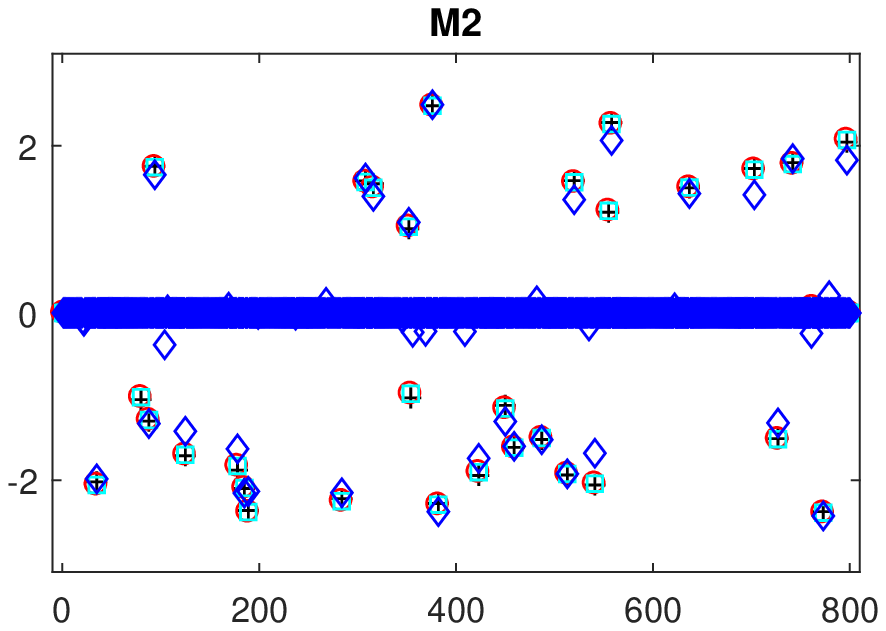}

\includegraphics[width=0.48\textwidth]{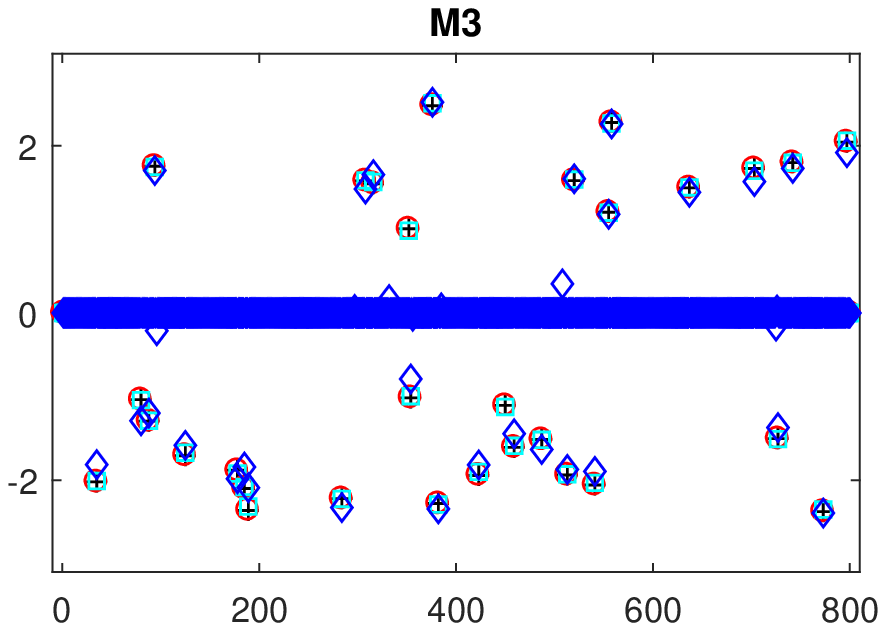}
\includegraphics[width=0.48\textwidth]{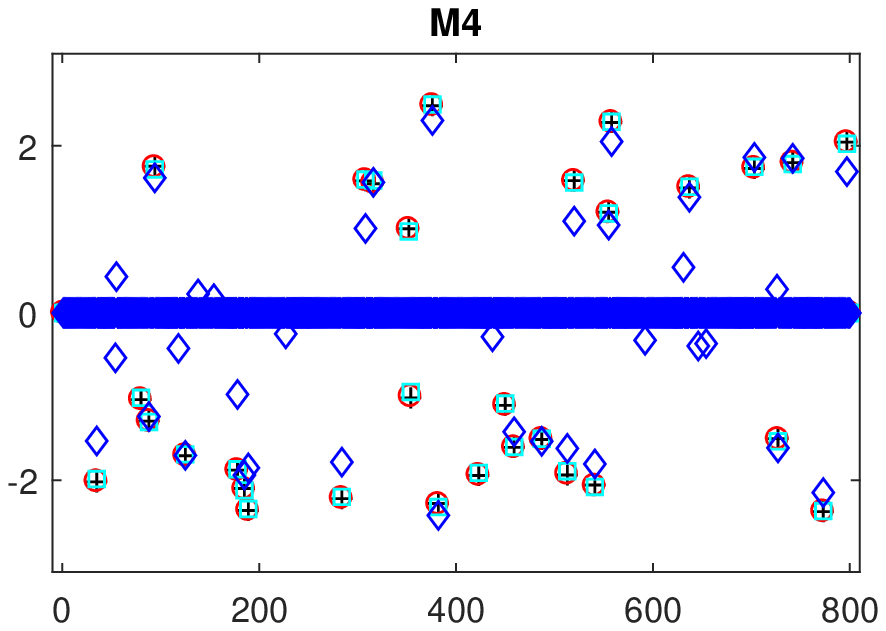}

\includegraphics[width=0.6\textwidth]{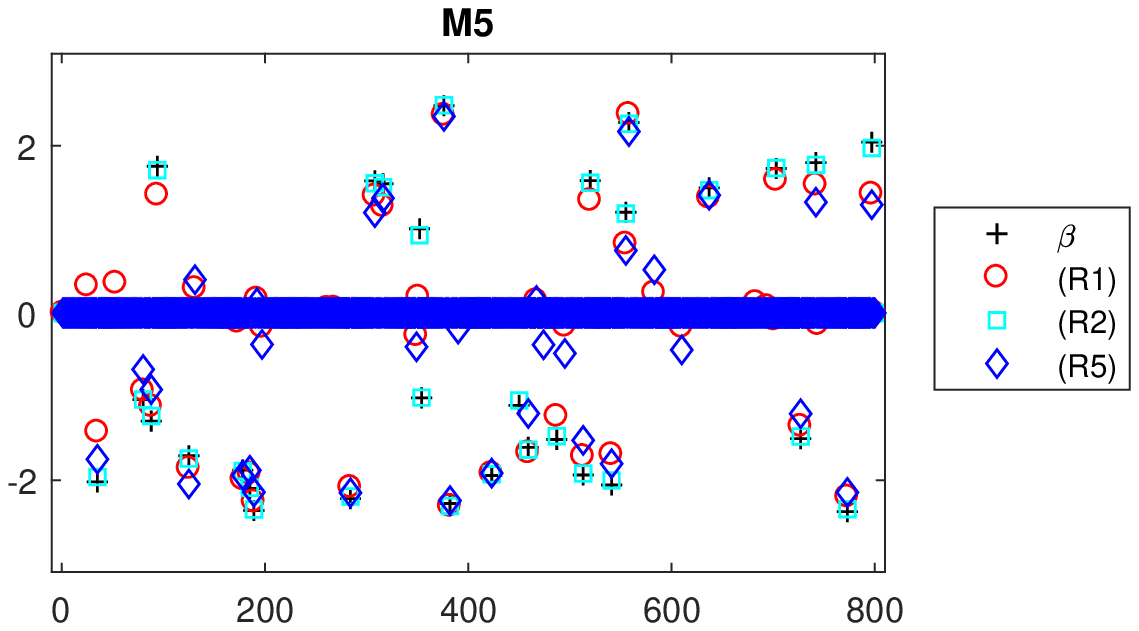}
\caption{An instance of the exact sparse vector $\beta$ and its recovery using different methods to generate $X$ and $\rho$.  Each figure indicates a different method (M1) - (M5) to generate $X$.  Within each figure, the symbols correspond to the different methods of generating $\rho$, with ${\color{red}\circ}$ the values of $\widehat{\beta}$ found using $\rho$ from (R1), ${\color{cyan}\square}$ using (R2), and ${\color{blue}\diamond}$ using (R5), where $+$ are the original values of $\beta$. }
\label{fig:CS}
\end{figure}

The results of the simulations are shown in Table~\ref{tab:CS} and Figure~\ref{fig:CS}.    The first two columns of the Table indicate the methods used to generate $X$ and $\rho$.  The third column is the average value of $\rho$ over 20 simulations for each pair (Mj,Rk).  The third column is the average of 20 simulations of the mean squared error of the ideal estimator~\cite{}, computed via
\begin{equation*}
	\mathrm{MSE} = \left( \frac{\left\| \beta - \widehat{\beta}\right\|_2^2}{\sum_{j=1}^{800} \min \{\beta_j^2, \sigma^2\}} \right)^{1/2}.
\end{equation*}
A smaller MSE indicates more accurate results. 
The final column is the interval estimates the restricted isometry constants, where $a = 1-\delta_\ell$ and $b = 1+\delta_r$ from Equation~\eqref{eq:RIP CS}, approximated similar to~\eqref{eq:RIP est} but restricting the sampling to $30$-sparse vectors.  

Figure~\ref{fig:CS} displays one instance of the values $\beta$ and $\widehat{\beta}$ for all pairs (Mj,Rk).  Note that (R1) and (R2) produce good results for any (Mj), except for (M5,R1), but (R3) largely gives unacceptable results.

%

\begin{table}[htpb]
\centering
\begin{tabular}{c c || c | c | c }
\multicolumn{2}{c||}{Methods} 	&$\rho$ 	&MSE 	 &$[a,b]$ \\
\hline
M1    &R1    &1.00   &1.34   &[0.82, 1.19]\\
M1    &R2    &1.00   &1.34   &[0.81, 1.19]\\
M1    &R4    &0.34   &20.12   &[0.28, 0.40]\\
\hline
M2    &R1    &1.00   &1.27   &[0.81, 1.19]\\
M2    &R2    &1.00   &1.27   &[0.81, 1.20]\\
M2    &R4    &0.34   &24.62   &[0.27, 0.40]\\
\hline
M3    &R1    &1.00   &{\color{red}0.14}   &{\color{red}[6.64, 9.71]}\\
M3    &R2    &0.12   &1.29   &[0.81, 1.19]\\
M3    &R4    &0.04   &21.19   &[0.27, 0.40]\\
\hline
M4    &R1    &1.00   &{\color{red}0.31}   &{\color{red}[2.89, 4.43]}\\
M4    &R2    &0.27   &1.41   &[0.79, 1.22]\\
M4    &R4    &0.09   &27.29   &[0.26, 0.41]\\
\hline
M5    &R1    &1.00   &13.74   &[0.35, 0.55]\\
M5    &R2    &2.23   &1.46   &[0.79, 1.24]\\
M5    &R4    &0.74   &28.93   &[0.26, 0.41]\\
\end{tabular}
\caption{Results of the sparse vector recovery simulations. The values highlighted in red correspond to the accuracy and restricted isometry intervals with the lowest MSE among all the simulations.}
\label{tab:CS}
\end{table}

Good results may be obtained using any method (M1)-(M5) to generate $W$.  If the restricted isometry interval is relatively tight and centered about 1, then the MSE is small.  This is to be expected, as good performance is guaranteed with high probability for matrices satisfying the RIP.  This is further evidenced by the results for (R2) when $\rho$ is scaled so all (Mj) behave as (M1) or (M2) with $\rho = 1$, which are proven to perform well in compressed sensing tasks. 
However, by far the best accuracy results are achieved for (M3, R1) and (M4, R1).  These methods are not generally used with compressed sensing, with no probabilistic guarantees for performance, and have seemingly undesirable restricted isometry intervals.

\section{CONCLUSIONS}\label{sec:conclusion}
The restricted isometry property gives guarantees with high probability of good performance of a compressed sensing scheme that is input data agnostic.  We adapted the RIP for random matrices arising in the hidden layer of echo state networks, with experimental results suggesting that matrices that have good near-isometry constants also give good results in ESN classification tasks.  The specific method used to generate the random behavior of $W$ is less important than the definition of the scaling factor $\rho$, which seems to perform consistently well using method (R2) with any random $W$.  Better performance may be attained using other approaches, as illustrated in the compressed sensing results in Table~\ref{tab:CS}, however these approaches do not have theoretical performance guarantees.  Further research into these methods may be warranted.

The results also suggest that the guideline of selecting $W$ and $\rho$ to satisfy $|\rho W|_*<1$ in an ESN for good performance is not necessarily a good strategy. We propose that a better, and more consistent approach, would be to select $W$ and $\rho$ to achieve the tightest near-isometry interval centered about 1.


\bibliographystyle{plain}
\bibliography{mybibfileRIP} 

\end{document}